\documentclass[aps,amssymb,prl,twocolumn,showpacs]{revtex4}
\usepackage{graphicx}
\usepackage{epsfig,amsmath}
\begin{document}

\title{Electrical manipulation and measurement of spin properties of quantum spin Hall edge states}
\author{Jukka I. V\"ayrynen}
\author{Teemu Ojanen}\email[Correspondence to ]{teemuo@boojum.hut.fi}
\affiliation{Low Temperature Laboratory, Aalto University, P.~O.~Box 15100,
FI-00076 AALTO, Finland }

\date{\today}
\begin{abstract}
We study effects of a gate-controlled Rashba spin-orbit coupling to quantum spin-Hall edge states in HgTe quantum wells. A uniform Rashba coupling can be employed in tuning the spin orientation of the edge states while preserving the time-reversal symmetry. We introduce a sample geometry where the Rashba coupling can be used in probing helicity by purely electrical means without requiring spin detection, application of magnetic materials or magnetic fields. In the considered setup a tilt of the spin orientation with respect to the normal of the sample leads to a reduction in the two-terminal conductance with current-voltage characteristics and temperature dependence typical of Luttinger liquid constrictions.
\end{abstract}
 \bigskip
\maketitle

The quantum spin Hall (QSH) insulator is a two-dimensional (2D) example of recently discovered topological insulators \cite{kane1,bernevig1,kane2}. Due to their special band structure arising from a strong spin-orbit interaction (SOI), topological insulators exhibit gapless surface modes forming a helical electron liquid. In the QSH state the edge supports two counter-propagating modes of opposite spin. The edge modes are protected by time-reversal symmetry, thus being robust against effects of small deformations of the sample and time-reversal invariant disorder \cite{wu}. The existence of the QSH state has been confirmed in a series of experiments performed in HgTe quantum well structures \cite{konig1,konig2,konig3,roth}. However, a quantitative observation of the helical structure has not been achieved yet.

Previously helicity detection has been considered in hybrid spin Hall-QSH structures where gaining a quantitative results has proven to be difficult \cite{konig3}. Hou and coworkers proposed a four-terminal geometry where the helical nature of the edge states manifest in a novel conducting phase \cite{hou}. Their scheme requires strong electron-electron interactions that can be achieved only under special conditions. Recently it was also proposed that injection of electrons from a ferromagnetic STM tip to the edge modes could reveal the helical nature of the edges \cite{das}. However, a purely electrical measurement is desirable since injection of a spin-polarized current from a ferromagnetic material is highly inefficient \cite{schmidt}.

\begin{figure}[h]

\centering
\includegraphics[width=0.63\columnwidth,clip]{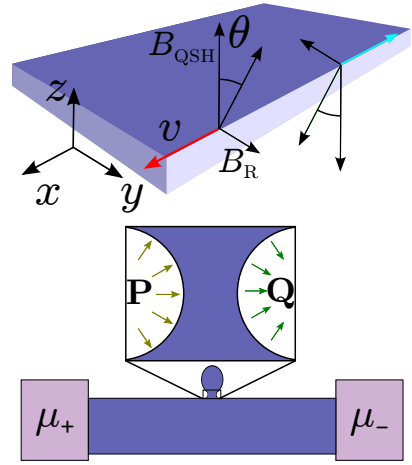}
\caption{Top: Left-moving and right-moving edge channels experience opposite effective magnetic fields $B_{\mathrm{QSH}}$ and $B_\mathrm{R}$ corresponding to the intrinsic QSH field and the Rashba SOI. The spin orientation of the edge states is determined by a relative magnitude of the fields. The effective field $B_R$ and thus the angle $\theta$ can be tuned by the external gate voltage.
Bottom: Quantum well with a smooth edge deformation (exaggerated in the figure) seen from above. The Rashba field induces an in-plane spin projection which is parallel transported along the edge. The magnified view shows the in-plane spin projection of right movers at $P$ and left movers at Q. Due to the rotation along the deformation, the spin states of the counter-propagating channels on different sides of the point contact have a finite overlap. This enables a spin-dependent tunneling which can be tuned by the gate voltage.}\label{scheme}
\end{figure}

The intrinsic SOI in HgTe quantum-well structures exceeding the critical thickness $d_c\approx 63 \,\mathrm{\AA}$ results in an inverted band structure leading to the QSH state \cite{bernevig1}. If, in addition, the surface inversion symmetry is broken by an asymmetric doping or application of a gate-induced electric field, also the Rashba SOI is present \cite{winkler}. The tunability of the Rashba interaction renders it a very attractive possibility for spin manipulation. The magnitude of the Rashba SOI in HgTe quantum wells can be several times larger than in any other semiconductor material and can be tuned over a wide range by varying the gate voltage \cite{hinz,konig4}. Recently, effects of a disordered Rashba interaction on edge states were studied in presence of electron-electron interactions. This type of disorder can lead to a localization and destroy the QSH state for sufficiently strong interactions \cite{strom1}.

In this paper we study an electronic manipulation and measurement of helicity properties of the edge states by employing the Rashba SOI. A spatially uniform Rashba SOI, controllable by the gate voltage, rotates the spin axis of the edge modes as depicted in Fig.~\ref{scheme} (top). We introduce a point-contact geometry where helicity of the edge modes can be probed by an electrical conductance measurement, Fig.~\ref{scheme} (bottom). The spin orientation of the edge modes manifests in a reduced two-terminal conductance with a distinctive current-voltage characteristics and temperature dependence typical for Luttinger liquid constrictions. This is our central result.

First we discuss the QSH edge model in the presence of the Rashba SOI and electron interactions. Then we introduce a point-contact geometry suitable for probing spin of the edge states and conclude by discussing experimental observation of predicted effects.

Low-energy dynamics of HgTe/CdTe wells are well-described by $k\cdot p$ -Hamiltonian derived in Ref.~\cite{bernevig1}. This description is applicable when studying phenomena on length scales much larger than the lattice constant. By imposing a boundary and projecting the 2D Hamiltonian onto the edge states (see Refs.~\cite{konig2}), one obtains the effective 1D Hamiltonian of the gapless edge excitations in the quantum spin Hall phase
\begin{equation}\label{basic}
H_{\mathrm{QSH}}=v_F\int dx\Psi^\dagger(-i\partial_x\sigma_z)\Psi,
\end{equation}
where the two-component spinor $\Psi=(\psi_{\uparrow},\psi_{\downarrow})^T$ describes the helical edge modes of a single edge ($x$ and $z$-directions as in Fig.~1). The edge states reflect the topological order in the bulk and are robust against deformations. The crucial property of the Hamiltonian (\ref{basic}) is that the counter-propagating modes consist of fermions of opposite spins. The $z$-direction coincides with the perpendicular direction of the 2D structure. In addition of (\ref{basic}) we consider a Rashba spin-orbit coupling of the form
\begin{equation}\label{rashba}
H_R=\int dx\alpha\Psi^\dagger(-i\partial_x\sigma_y)\Psi,
\end{equation}
where $\alpha$ is the spin-orbit coupling constant that can be tuned by the external gate voltage. An interaction of the form (\ref{rashba}) arises from the surface inversion asymmetry induced by the perpendicular electric field. For the remainder of our analysis we treat $\alpha$ as a real constant independent of position. The Rashba term (\ref{rashba}) is also time-reversal invariant so the fundamental symmetries of the system are respected. Thus, in the absence of electron-electron interactions, the edge states are described by the Hamiltonian  $H_0=H_{\mathrm{QSH}}+H_R$
\begin{equation}
H_0=v_\alpha\int dx\Psi^\dagger(\mathrm{cos}\,\theta\sigma_z+\mathrm{sin}\,\theta\sigma_y)(-i\partial_x)\Psi,\nonumber
\end{equation}
where $v_\alpha=\sqrt{v_F^2+\,\alpha^2}$, $\mathrm{cos}\,\theta=v_F/v_\alpha$ and $\mathrm{sin}\,\theta=\alpha/v_\alpha$. Introducing the rotation $\Psi'=\mathrm{exp}(-i\sigma_x\theta/2)\Psi$, $H_0$ can be brought in the form (\ref{basic}):
\begin{equation}\label{zero}
H_0=v_\alpha\int dx\Psi'^{\dagger}(-i\partial_x\sigma_{z'})\Psi',
\end{equation}
where the new $z'$-direction forms an angle $\theta$ with the $z$-axis. In the following we will drop the primes from the rotated quantities in Eq.~(\ref{zero}) remembering that the new $z$-axis is rotated with respect to the normal of the plane, as shown in Fig.~\ref{scheme}. The Hamiltonian (\ref{zero}) indicates that the uniform Rashba term results in a renormalization of the Fermi velocity and a rotation of the spin axis, leaving the system otherwise unaffected. However, the tilt in the spin axis leads to qualitatively new effects when the edge geometry has deformations such as circular arcs. The in-plane component of spin, tunable by the gate voltage, follows along the edge as depicted in Fig.~\ref{scheme}. This property can be used in detection of the helicity direction by a purely electrical measurement as discussed below.

The Hamiltonian (\ref{zero}) describes the edge channel of noninteracting particles and provides a natural starting point to discuss interactions. Only interactions allowed by the time-reversal invariance are of the form \cite{wu}
\begin{equation}
H_d=g_d\int\,dx\psi^\dagger_{\uparrow}\psi_{\uparrow}\psi^\dagger_{\downarrow}\psi_{\downarrow},\,\, H_f=g_f\int dx\psi^\dagger_{\sigma}\psi_{\sigma}\psi^\dagger_{\sigma}\psi_{\sigma},\nonumber
\end{equation}
where the first term corresponds to density-density interaction between the counter-propagating modes and the second term describes forward scattering (index $\sigma$ is summed over the two spin projections). There is one more allowed interaction term describing momentum non-conserving umklapp processes which is unimportant away from the half filling and not considered here. By employing standard bosonization methods \cite{giamarchi}, the interacting Hamiltonian $H=H_0+H_d+H_f$ becomes
\begin{align*}
H=\dfrac{u}{2\pi}\int\, dx\left(g(\partial_{x}\theta)^{2}+g^{-1}(\partial_{x}\varphi)^{2}\right),
\end{align*}
where $g=\sqrt{\dfrac{2\pi v_{\alpha}+g_{f}-2g_{d}}{2\pi v_{\alpha}+g_{f}+2g_{d}}}$
is the Luttinger liquid interaction parameter and $u=v_{\alpha}/g$ the renormalized Fermi
velocity.

\begin{figure}[h]
\centering
\includegraphics[width=0.4\columnwidth,clip]{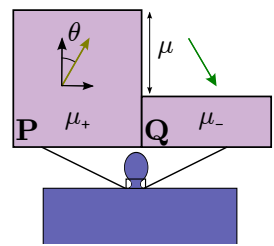}
\caption{Right-moving and left-moving edge channels on different sides of the point contact (P and Q) have different chemical potentials reflecting the bias voltage over the system. Due to the finite tilt $\theta$, the spin states of the counter-propagating electrons at P and Q have finite overlap (proportional to $\sin\theta)$ enabling tunneling. Tunneling current reduces the total current propagating along the edge state. By tuning $\theta$ it is possible to tune the backscattering current through the point contact.
}\label{point}
\end{figure}

Now we examine consequences of the Rashba SOI induced tilt in the two-terminal transport setup depicted in Fig.~\ref{scheme}. As indicated in Fig.~\ref{point}, the counter propagating electrons on the different sides of the point contact have a potential difference due to the applied bias between the sample edges. In addition, the rotation of the in-plane spin component along the arc enables the tunneling of right movers to left movers (and vice versa when the bias is inverted) through the point contact, thus reducing the total current propagating along the edge. To couple the edge states on different sides, the width of the constriction should be of the order of 200 nm or smaller \cite{zhou}. Samples of comparable dimensions have already been fabricated and employed in experiments \cite{bruene}. Calculation of the tunneling current is analogous to the analysis of Ref.~\onlinecite{wen}.
The tunneling Hamiltonian through the point contact, with potential difference $\mu=eV>0$ between the
right- and left-moving electrons, is $
H_{\text{tun}}(t)=\Gamma_{+}(t)\psi_{-}^{\dagger}(Q)\psi_{+}(P)+\text{h.c.}+Q\leftrightarrow P$
where $\Gamma_{+}(t)=\sqrt{\mathcal{T}}e^{i\mu t}\sin\theta $
is the effective tunneling amplitude. The factor $e^{i\mu t}$ appears when the potential difference is accommodated through the Peierls substitution and $\sin\theta$ arises from the overlap of the right-moving and left-moving spin states on different sides of the contact. We have labeled the electron field operators as right ($\psi_{+}$) and left ($\psi_{-}$) movers bearing in mind that the spin-orientation is bound to the propagation direction.
The tunneling current operator is the time-derivative of the number
operator: $\dot{N}_{+}(t)=i\int dx'\left[H_{\text{tun}}(t),\psi_{+}^{\dagger}(t,x')\psi_{+}(t,x')\right]=i\Gamma_{+}(t)B(t)+\text{h.c.},$
where $B(t)=\psi_{-}^{\dagger}(t,Q)\psi_{+}(t,P)+Q\leftrightarrow P$.
The expectation value of the current through the point-contact
in the lowest order of the tunneling becomes $I_{\mathrm{pc}}=-2e\mathcal{T}\sin^{2}\theta\,\text{Re}\int dt\,\Theta(t)e^{i\mu t}\left\langle \left[B(t),B^{\dagger}(0)\right]\right\rangle $
where the expectation value can be expressed in terms of Green's function $-2i\,\text{Im }G_{+}^{>}(t,0)^{2}$
calculated in the absence of $H_{\text{tun}}(t)$. For positive times we get \begin{align*}
G_{\pm}^{>}(t,0)= & i^{-2\gamma}\dfrac{D^{-2\gamma}}{2\pi u}\left\{ \dfrac{\beta}{\pi}\sinh\dfrac{\pi}{\beta}(t-i0^{+})\right\} ^{-2\gamma-1}\end{align*}
where $D=u/a$, $a$ is the short-distance cutoff and $\gamma=(g+g^{-1})/4-1/2$.
Thus we have $
I_{\mathrm{pc}}=-4e\mathcal{T}\sin^{2}\theta\,\text{Im }X(\mu)$,
with the Fourier sine-transform of the response function $X(\mu)=\int\mathrm{d}t\,\Theta(t)\sin\mu t\, G_{+}^{>}(t,0)^{2}.$
Evaluation of the integral yields
\begin{flalign}
I_{\mathrm{pc}}= & -4e\mathcal{T}\sin^{2}\theta\,\dfrac{D^{-4\gamma}}{(2\pi u)^{2}}\left(\dfrac{2\pi}{\beta}\right)^{4\gamma+1}\sinh\dfrac{\beta eV}{2}\nonumber \\
 & \times B(2\gamma+1-\dfrac{i\beta eV}{2\pi},2\gamma+1+\dfrac{i\beta eV}{2\pi})\label{eq:finiteTcurrent}
\end{flalign}
in terms of the the Euler beta function $B(x,y)$. Here we have implicitly assumed that the interference between the two backscattering channels which correspond to electrons traversing the arc in opposite directions is suppressed \footnote{ This is justified when the loop is longer than the dephasing length $L_0$, where the dephasing length has a typical Luttinger liquid expression $L_0=u/4\pi\gamma T$. If the arc is shorter that the $L_0$ it is necessary to include the interference term: in a limit where the arc length is much shorter than $L_0$ the two paths interfere destructively and the backscattering current through the point contact vanishes.}.
From the current we can evaluate the zero-bias conductance as 
\begin{align}
G_{\mathrm{pc}}=2e^{2}\mathcal{T}\sin^{2}\theta\,\dfrac{D^{-4\gamma}}{2\pi u^{2}}\left(\dfrac{2\pi}{\beta}\right)^{4\gamma}B(2\gamma+1,2\gamma+1).\nonumber
\end{align}
At zero temperature the current-voltage relation becomes \[
I_{\mathrm{pc}}=-2e\mathcal{T}\sin^{2}\theta\,\dfrac{D^{-4\gamma}}{2\pi u^{2}}\dfrac{(eV)^{4\gamma+1}}{\Gamma(4\gamma+2)}.\] which is not directly relevant since our assumptions concerning backscattering do not hold at $T=0$. The current-voltage relation exhibits an interaction-dependent power law typical for Luttinger liquids, reflecting the anomalous dimension $(g+g^{-1})/2$ of the tunneling operator \cite{strom2}.
\begin{figure}[h]
\centering
\includegraphics[width=1.05\columnwidth,clip]{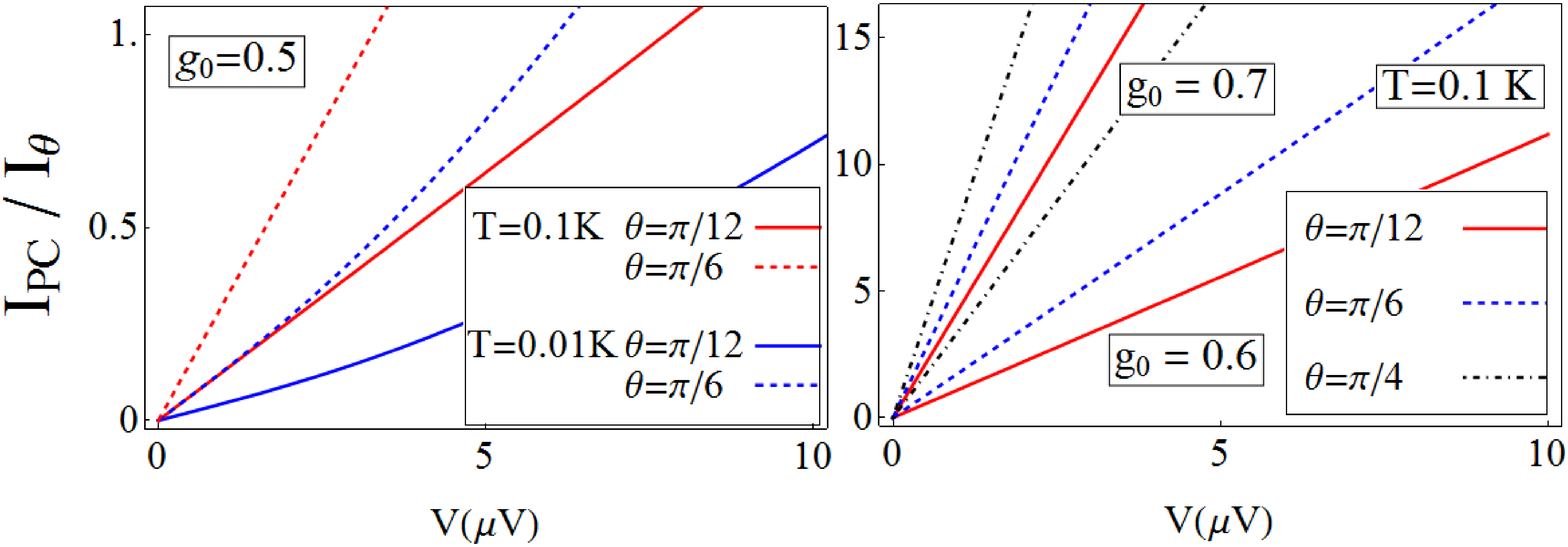}
\caption{I-V characteristics of the point-contact current at different temperatures and interaction strengths.}\label{IV}
\end{figure}
\begin{figure}[h]
\centering
\includegraphics[width=0.64\columnwidth,clip]{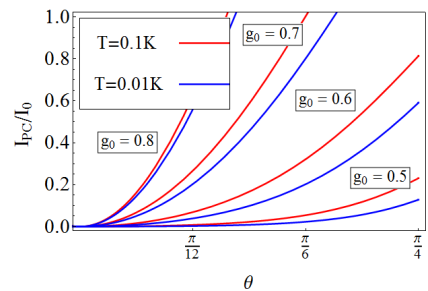}
\caption{Angular dependence of the point-contact current corresponding to different interaction strengths, where $g_0$ denotes the value of $g$ at $\theta=0$. $I_{\mathrm{pc}}$ is a symmetric function of $\theta$. The curves represent the bias value $V=1\, \mu$V}\label{modul}
\end{figure}
In addition to the explicit $\theta$-dependence, the point-contact current (\ref{eq:finiteTcurrent}) depends on $\theta$ through $u$ and $g$ since the effective Fermi velocity $v_{\alpha}$ is $\theta$-dependent. To study the angular dependence of $I_{\mathrm{pc}}$, it is convenient to decompose it as $I_{\mathrm{pc}}=I_{\theta}F(V,T,\theta)$, where the prefactor $I_\theta=\frac{4e\mathcal{T}\times10^{-7}}{av_F}\sin^2\theta\cos^{4\gamma+2}\theta$ encodes the primary $\theta$-dependence and $F(V,T,\theta)=I_{\mathrm{pc}}/I_{\theta}$ is a dimensionless function plotted in Fig.~\ref{IV} \footnote{The factor $10^{-7}$ appearing explicitly in $I_{\theta}$ arises from normalization of $F(V,T,\theta)$ to the order of unity in the plotted regime.}. As is evident, also $F(V,T,\theta)$ exhibits a moderate $\theta$-dependence. A typical value of the interaction parameter $g_0$ in HgTe quantum wells in the absence of the Rashba SOI has been estimated as $g_0\gtrsim 0.5$ \cite{hou,strom1}. To estimate the typical magnitude of the current $I_0=\frac{4e\mathcal{T}\times10^{-7}}{av_F}$ we assume realistic values $v_F=5.5\times 10^5$ m/s, $a=10$ nm and $\mathcal{T}=0.1v_F^2$, implying that $I_0=4\times10^{-3}$ nA. These numerical values suggest that $G_{\mathrm{pc}}=0.01-0.1\times 2e^2/h$ is achievable.

The current through the point contact enables spin-dependent backscattering, reducing the two-terminal current and conductance from the values $I$, $G$ in the absence of the Rashba term. The total current between the terminals becomes $|I_\mathrm{t}|=|I|-2|I_{\mathrm{pc}}|$ implying that the conductance is reduced by the point-contact contribution $G_\mathrm{t}(T,\theta)=G(T)-2\,G_{\mathrm{pc}}(T,\theta)$. The magnitude of $I_{\mathrm{pc}}$ is a symmetric function of $\theta$ and grows rapidly away from $\theta=0$, see Fig.~\ref{modul}. By varying $\theta$ through the gate voltage and measuring $G$ one can access $G_{\mathrm{pc}}(T,\theta)$.

The tunability of the Rashba SOI is crucial for observing the spin-dependent conductance. The effective Rashba field should, ideally, be comparable to the intrinsic QSH field to reach significant tilt angles. Using the parameters calculated in Ref.~\onlinecite{rothe}, we estimate that the Rashba coupling for a quantum well of width $70\,\mathrm{\AA}$ is $\alpha=-15.6\,\mathrm{nm}^2e\mathcal{E}_z $, where $\mathcal{E}_z$ is a perpendicular electric field. In case the gate voltage between the well and the gate electrode 100 nm apart is 1 V, the estimate yields  $\alpha\approx 0.16$ nm eV \footnote{There also exists Rashba terms quadratic and cubic in $k$ \cite{rothe}. For small $k_F\sim 0.1$ nm$^{-1}$ of the edge states these terms are less than ten and one percent of the linear term.}. This value should be compared to the QSH scale $A=\hbar v_F=0.365$ nm eV, so for realistic gate voltages the Rashba SOI and the intrinsic QSH field are of the same order. This suggests that even large tilt angles $\theta\approx\pi/4$ are achievable. Tuning the gate voltage does not only affect the Rashba interaction but in practice also changes the Fermi energy of the system. This could be undesired since at large gate voltages the system moves from the QSH state to the metallic regime. The effect can be compensated by application of a back gate which allows tuning of the Rashba SOI and the Fermi energy independently. Experimental efforts towards realizing these type of structures are already underway.

Another challenge is measuring the backscattering current due to the point contact. In real samples there are always unideal features reducing the theoretical conductance value $G_Q=2e^2/h$. The residual scattering, attributed to local conducting regions close to the edges \cite{roth}, can be distinguished from the point-contact backscattering due to their different dependence on the gate voltage and temperature. As long as the system stays in the QSH regime, changes in the gate voltage are not expected to affect the conductance essentially. Especially, if only the asymmetry of the well is changed keeping the Fermi level constant, residual backscattering should remain constant. However, the point-contact current is strongly gate dependent and vanishes at the gate value corresponding to a symmetric well. The maximum conductance is obtained at this gate value and any departures from the symmetric configuration leads to reduction enabled by the tilted spin orientation. In addition, as long as the temperature of the system is smaller than the bulk energy gap, the QSH state should not be affected by increase of temperature. In contrast, the point-contact current leads to a significant reduction of conductance when temperature is increased. By measuring conductance as a function of the gate voltage and temperature it is possible to resolve the point-contact contribution.

In summary, we proposed a purely electrical manipulation and characterization of spin properties of helical QSH edge states in HgTe quantum wells. A gate-controlled electric field induces a finite Rashba SOI which can be employed in tuning the spin orientation of the edge states. This results in a tunable in-plane spin projection following smooth deformations of the edges. We introduced a point-contact geometry where helicity can be probed by a spin-dependent backscattering accessible in two-terminal conductance measurements.

The authors are deeply indebted to Markus K\"onig for invaluable advice and comments. The authors acknowledge support from the Academy of Finland (T.O) and ERC  Grant No. 240362-Heattronics (J.I.V).

\end{document}